# Atom-by-Atom Direct Writing

*Ondrej Dyck[1], Andrew R. Lupini[1], Stephen Jesse[1]*

*[1] Center for Nanophase Materials Sciences, Oak Ridge National Laboratory, Oak Ridge, TN*

## Abstract

Direct-write processes enable the alteration or deposition of materials in a continuous, directable, sequential fashion. In this work we demonstrate an electron beam direct-write process in an aberration-corrected scanning transmission electron microscope. This process has several fundamental differences from conventional electron beam induced deposition techniques, where the electron beam dissociates precursor gases into chemically reactive products that bond to a substrate. Here, we use elemental tin (Sn) as a precursor and employ a different mechanism to facilitate deposition. The atomic-sized electron beam is used to generate chemically reactive point defects at desired locations in a graphene substrate. Temperature control of the sample is used to enable the precursor atoms to migrate across the surface and bond to the defect sites thereby enabling atom-by-atom direct-writing.

The scanning transmission electron microscope (STEM) has a long legacy as an indispensable materials characterization platform. In recent years this picture has begun to expand to include not only characterization but also manipulation.[1–3] From this perspective the STEM can be viewed as a fabrication platform capable of addressing individual atoms with the highly focused electron beam (e-beam) and altering materials atom-by-atom. Such capabilities hearken back to Feynman's famous quote about putting every atom where we want.[4] One would like to be able to build structures from the atom up, designed for their electronic, optical, chemical, or structural properties. Quantum qubits, quantum sensors, and perhaps even full computing circuits could be tailored with atomic specificity to elicit the emergent quantum properties of the materials. Creating structures with atomic specificity would throw open the doors for experiments developing fundamental understanding of physics, chemistry, and materials science at the atomic level. Realizing such a vision, however, is a more difficult task.

Observations of e-beam induced sample alterations date back to some of the earliest days of electron microscopy. Every electron microscopist is familiar with e-beam damage[5–7] and hydrocarbon deposition. Mitigation of these unintentional processes has been a key enabling factor for the (S)TEM to excel at materials characterization. Lower accelerating voltages,[8,9] enhanced by aberration correction,[10] have been effective at reducing knock-on damage. With unintended processes at least partially under control, we can begin to isolate variables to examine their effects on the specimen with a view toward reproducible atomistic transformations.

To simplify matters further, here we work with 2D materials where, instead of looking at projected columns of atoms (as in a 3D crystal), each atom can be uniquely identified in the image. The most robust and well-studied 2D material is graphene and many studies have been carried out to find ways to clean and image contamination-free graphene.[11–24] With these advancements, which act to mitigate major alterations to the specimen, minor alterations at the atomic scale can be observed and studied. Some examples of these atomic scale dynamics are shown in Figure 1(a)-(d): the spinning of a Si molecular rotor,[25] the dynamic rearrangement of a $Si_6$ cluster embedded in a graphene nanopore,[26] the movement of a 3-fold coordinated Si dopant in graphene, and vacancy diffusion in graphene.[27] In these examples the major focus was to observe and understand the e-beam driven effects. The transition to top-down control over atomic movement stemmed from the realization that the position of the e-beam could influence the direction of motion of the 3-fold coordinated Si atoms. The first demonstration of precisely directed movement of a single atom using an e-beam is shown in Figure 1(e) where a 3-fold coordinated Si atom was moved seven steps through a graphene lattice.[28] This result was independently replicated shortly afterward[29] and extended to longer distances,[30,31] shown in Figure 1(f) and (g). The first e-beam assembly of a primitive structure, a Si dimer, is shown in Figure 1(h).[30] Controlled movement of a Si atom through the wall of a single walled carbon nanotube is shown in Figure 1(i)[32] and intentional rotation of a Si trimer, as well as the reversible conversion to a tetrimer is shown in Figure 1(j).[30]

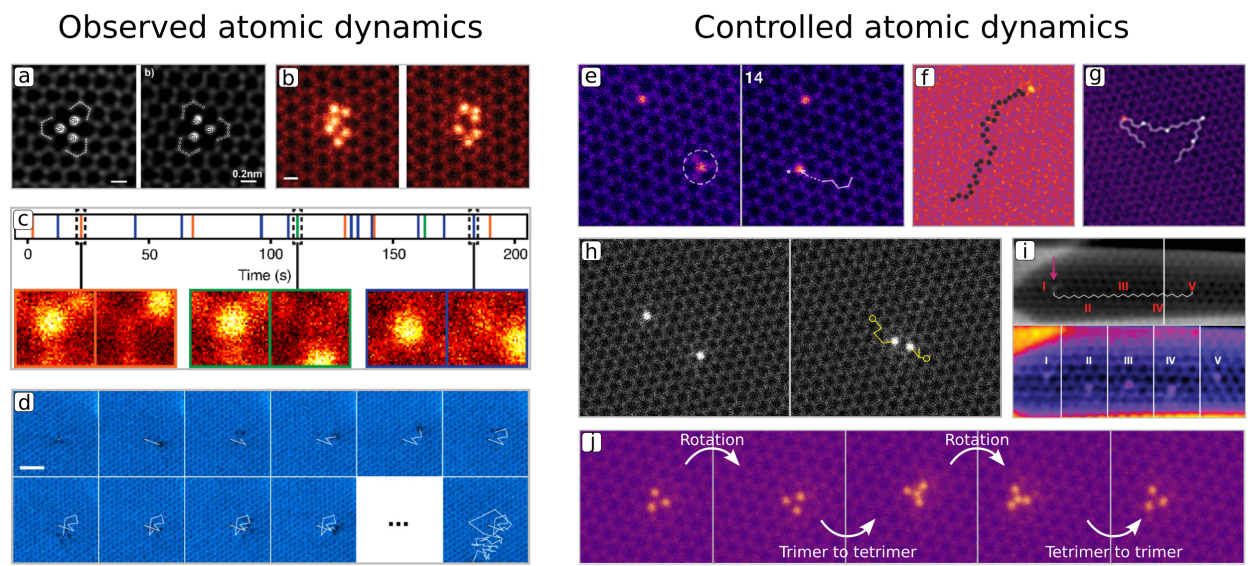

**Figure 1 Select examples of observed and controlled atomic dynamics in graphene.** (a) Observation of Si trimer rotation. (b) Observation of reversible transformation of a $Si_6$ cluster embedded in a graphene nanopore. (c) Observation of movement of 3-fold coordinated Si dopant in graphene. (d) Observation of vacancy diffusion in graphene. (e) Controlled movement of 3-fold coordinated Si dopant in graphene. (f) and (g) Controlled movement of 3-fold coordinated Si atom over an extended distance. (h) Controlled assembly of a Si dimer in graphene. (i) Controlled movement of Si atom in a single walled carbon nanotube. (j) Controlled rotation of a Si trimer and conversion to tetrimer. (a) Adapted from reference [25]. Reuse is permitted per the United States Department of Energy public access plan http://energy.gov/downloads/doe-public-access-plan. (b) Adapted with permission from reference [26]. (c) Reproduced with permission from reference [33] https://creativecommons.org/licenses/by/3.0/. (d) Reproduced with permission from reference [27] https://creativecommons.org/licenses/by/3.0/. (e) Adapted with permission from reference [34]. (f) Reproduced with permission from reference [30]. (g) Reproduced with permission from reference [31] https://pubs.acs.org/page/policy/authorchoice_ccby_termsofuse.html. (h) Adapted with permission from reference [30]. (i) Reproduced with permission from reference [32] https://creativecommons.org/licenses/by/4.0/. (j) Adapted with permission from reference [30].

While there is a certain elegance to the ability to move a single atom through the graphene lattice, it is not obvious that positioned Si atoms in graphene have applications aside from the development of these experimental techniques and unraveling the fundamental physics of beam-matter interactions. Likewise, the Si atom movement typically terminates with the ejection of an additional carbon atom converting it into the much less mobile 4-fold coordination.[31] This conversion occurs unpredictably, meaning that the distance a 3-fold coordinated Si atom can be moved is also unpredictable. First principles simulations of the assembly of the dimer shown in Figure 1(h) indicate that this assembly process is energetically favorable, which facilitates bringing atoms together. However, this energetic trend to agglomerate further complicates the predictability of atom movement in the presence of other dopants. Thus, the tailoring of atomic structures in this way is inherently limited.

Investigations have been undertaken to extend the types of dopant atoms than can be put in graphene. The main, serendipitous factor that facilitated the studies shown in Figure 1 was that Si dopant atoms are commonly found in graphene. Moving from serendipity to predictability, it was shown that the Si atoms in the surface contamination could be introduced into the graphene lattice by deliberately ejecting C atoms with the e-beam to create attachment sites for the Si atoms.[29,30] Dopant Si atoms could then be reliably produced using the e-beam as a manipulation tool. This technique was extended and applied generally to many elements where the source materials were first introduced onto the graphene surface.[35–37] Common to each of these demonstrations was the limitation that dopant insertion was only accomplished within a few nanometers from the parent nanoparticle and required a two-step procedure where the e-beam was used both for creating defects in the graphene substrate as well as separating source atoms from the parent nanoparticles. These and other related studies[21,28,32,38–50] suggest it might be possible to find appropriate conditions under which a continuous atom-by-atom direct-writing procedure may be realized. A direct-write mechanism would enable the creation of longer-range atomic structures without requiring the step-by-step movement of single atoms through the lattice. Here, we aim to demonstrate this possibility.

For an atom-by-atom direct-writing procedure to be of practical use we must address several challenges. Conceptually, a direct-write procedure binds a material to a substrate through some local chemical or physical influence. This conceptualization glosses over the atomistic details of the process in favor of the more abstract representation that more easily describes the intended goal. In a similar way, common descriptions of direct-write deposition such as electron beam induced deposition (EBID) do not spell out the details of the chemical attachment of every atom to the substrate. However, as we scale down a writing process, eventually we must confront the atomistic nature of the world and begin to address the fact that "binding a material to a substrate" is a vague statement. If we are to write with individual atoms, we must be specific about what we mean. Here, we mean that atoms, differing elementally from the substrate, form strong chemical bonds to defect sites introduced in the substrate through a localized influence. With this formulation we can begin to specify the requirements for writing at the atomic scale. We must have source materials (atoms) that are delivered to the region of interest. We must have a highly localized influence on the sample that is able to create chemically reactive sites with which the source atoms can bond. These bonds need to be strong enough to prevent spontaneous thermal diffusion over long time spans. The definition of "long", clearly, depends on the application but here we envision stability for days, ideally years, at ambient temperatures. Critically, we must have a clean surface on which to write, in order to provide a spatially uniform environment that enhances predictable control over the process.

To satisfy these constraints we use a suspended single layer graphene substrate, operate our microscope above the knock-on threshold for graphene, and thermally deliver Sn atoms to the damaged regions *in situ*. Atmospheric pressure chemical vapor deposition (AP-CVD) was used to grow graphene on Cu foil.[51] Poly(methyl methacrylate) (PMMA) was spin coated over the graphene for protection. A wet transfer[29] was used to transfer the graphene to a Protochips™ heater chip substrate and Sn was evaporated onto the graphene surface in a vacuum chamber using a Sn coated W filament. The graphene was pre-cleaned using a rapid thermal ramp to 1200

°C[23] and the Sn evaporation was carried out while the sample temperature was maintained at 500 °C.

A Nion UltraSTEM 200 operated at an accelerating voltage of 100 kV and nominal beam current of 17 pA was used to examine the sample. Prior to examination, the sample, holder cartridge, and magazine were baked in vacuum for eight hours at 160 °C. After introduction into the pole piece, the sample was again heated to 1200 °C for a few minutes to remove residual contaminants. The sample was cooled to 500 °C and held at that temperature for ~24 hrs. On returning the sample to room temperature, e-beam induced hydrocarbon deposition was observed, so the sample was heated again to 300 °C for further examination. No additional hydrocarbon deposition was subsequently observed at this temperature.

Figure 2(a) and (b) show two high angle annular dark field (HAADF) images of the graphene and Sn nanoparticles at 300 °C. Much of the sample was multilayer and the Sn nanoparticles attached predominantly at step edges. We note the presence of single Sn atoms decorating the step edges and various other locations. Figure 2(c) shows an example EELS spectrum acquired from one of the nanoparticles confirming that they are made of Sn.

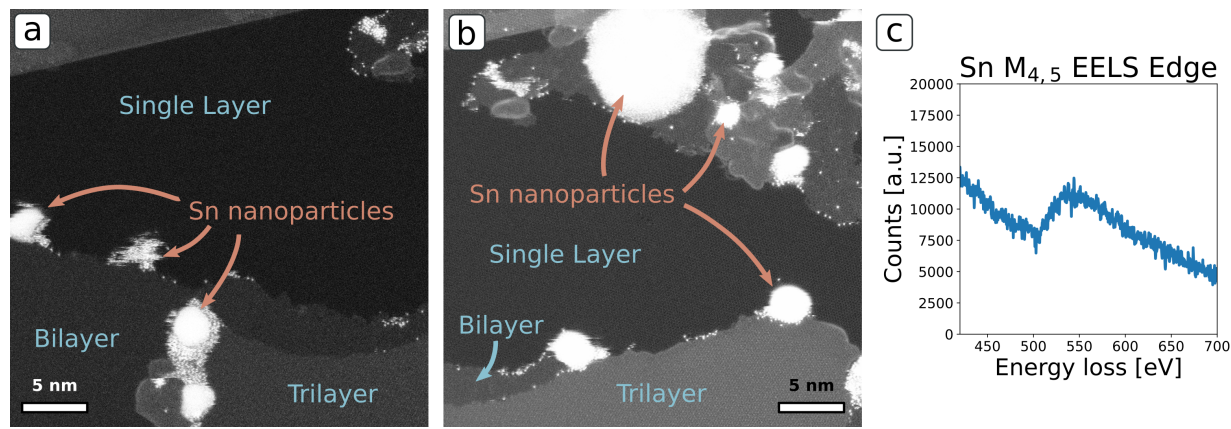

**Figure 2 Sample overview illustrating Sn nanoparticles attached to graphene.** (a), (b)

representative HAADF images of Sn nanoparticles attached to the graphene surface particularly at step edges. (c) Sn $M_{4,5}$ EELS edge acquired on one of the nanoparticles confirming they are composed of Sn.

Following the "beam dragging" strategy laid out in previous work,[30,35–37] Sn atoms were inserted into the graphene lattice with the substrate maintained at 300 °C. The e-beam was positioned on the source Sn particle and manually moved from the particle to the graphene, acting to both sputter Sn atoms from the parent nanoparticle and generate point defects in the graphene lattice for the Sn atoms to occupy. This procedure is depicted schematically in Figure 3(a)-(d). In the initial configuration, Figure 3(a), the e-beam is positioned on a nanoparticle of source material sitting on the graphene substrate. The e-beam is then moved across the substrate, depicted in Figure 3(b). This procedure generates vacancies in the graphene through direct ejection of lattice carbon atoms. The e-beam is then moved back to the source material and used to sputter atoms from the parent nanoparticle, Figure 3(c). The freed atoms can diffuse away from the nanoparticle across the graphene surface. There is some probability for vacancy diffusion and lattice restructuring; a detailed study and dynamic calculations are given in reference [52]. Figure 3(c) illustrates an example where two vacancies have merged forming a reconstructed divacancy. This structure does not have any dangling carbon bonds, which suggests that it (and other similar defect structures) will be less attractive for a diffusing adatom. Figure 3(d) depicts the capture of a diffusing atom by the remaining vacancy.

The generation of vacancies, sputtering of source atoms, and diffusion of vacancies and atoms are all stochastic processes, which means they are not guaranteed to occur. Typically, the e-beam must be repeatedly moved back and forth and the sample state checked via imaging after some time. Figure 3(e)-(g) shows a summary of the result of this procedure performed experimentally. Sn atoms are ejected from the parent nanoparticle and incorporated into the graphene lattice. As was also true in prior publications,[29,30,35–37] this process works across the short distance of a few nanometers and requires repeated attempts at defect creation and sputtering. This example was carried out with the substrate held at 300 °C and gives a reference point for the behavior of the graphene/Sn system which was qualitatively like the behavior of other elements at room temperature.[35–37] We would like to find conditions under which the stochastic nature of the dopant insertion process can be mitigated, so that dopant insertion can proceed in a continuous fashion. In addition, we would like to find conditions under which the dopant insertion process can be extended spatially and not remain confined to regions close to the parent nanoparticle.

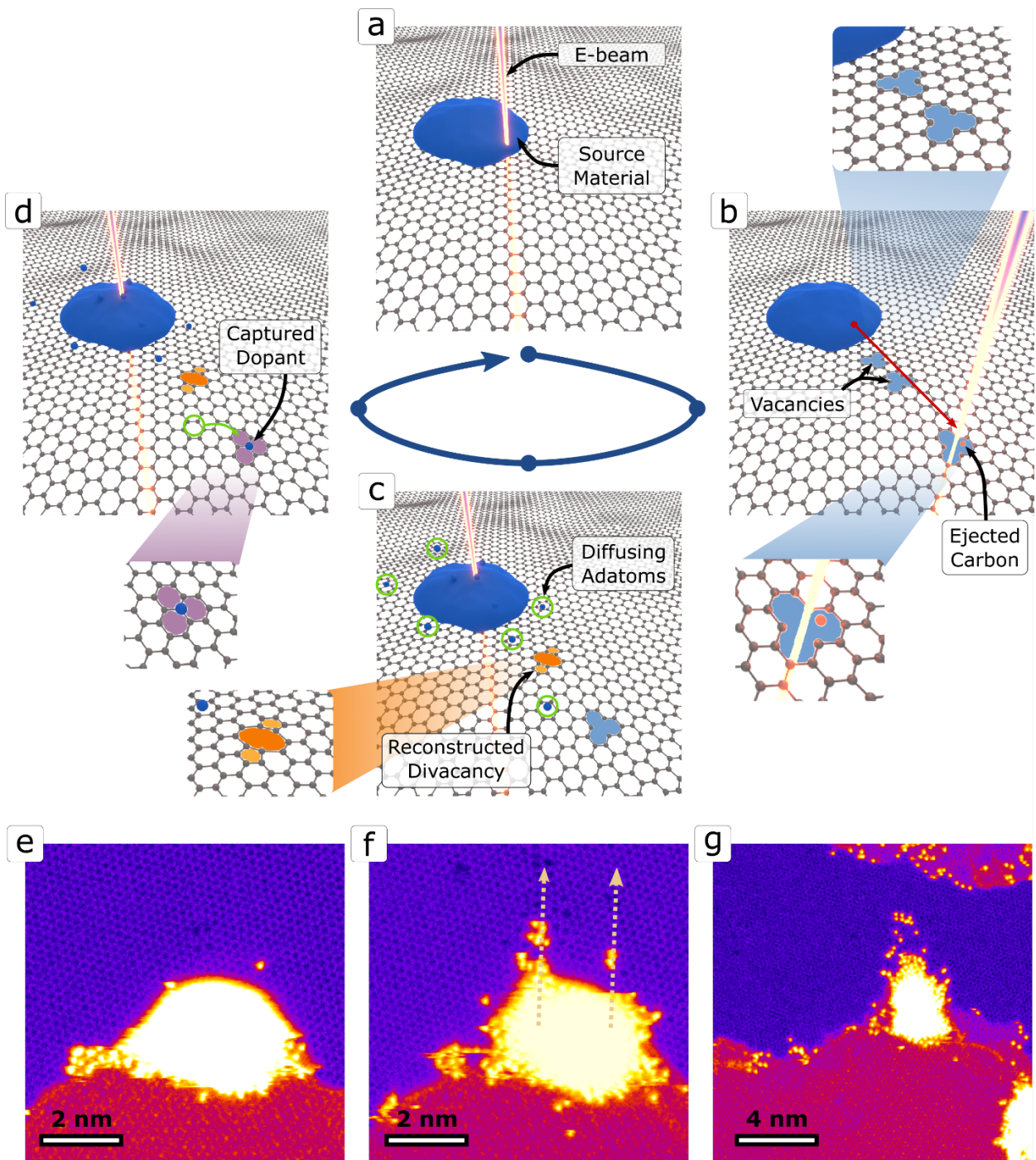

**Figure 3 Example of dopant insertion into single layer graphene.** (a) Schematic of e-beam positioned on source material. (b) e-beam is manually dragged across the substrate creating vacancies. (c) E-beam is repositioned on the source material and atoms are sputtered off. (d) adatoms can bond to the created vacancy sites. (e) Initial sample state where a Sn nanoparticle is attached to a step edge overlapping the single layer region. (f) "beam dragging" was performed in the locations indicated by the arrows resulting in Sn dopant insertion along the beam trajectory. (g) final state of the sample.

It is worth emphasizing that the e-beam in this and prior examples of dopant insertion is performing two distinct functions: 1) the generation of vacancies/defects in the graphene substrate to which the source atoms can chemically bond (Figure 3(b)), and 2) the ejection/sputtering of source atoms from the parent nanoparticle (Figure 3(c) and (d)). This means that the rate of supply of source atoms, freed from the parent nanoparticle, is dependent on e-beam current and position. More source atoms could likely be supplied by increasing the e-beam current, however, this also increases the defect generation rate. Because the defects are generated when the e-beam is positioned off the nanoparticle, the supply of source atoms is not continuously maintained. It is likely that these constraints are the main reason why dopant insertion with this method has only been accomplished within a few nanometers of the parent nanoparticle and why multiple attempts must be made before dopant insertion is accomplished.

To enhance our understanding of the processes described above quantitative modeling and measurement of the availability of source atoms as a function of e-beam current or position, for example, could be performed. However, the crux of the dilemma is that this process is reliant on the e-beam for both sputtering source material and the creation of defects. In contrast, e-beam induced deposition (EBID) does not rely on the e-beam to generate the flow of precursor material, and ideally the same principle should apply here. We seek appropriate sample conditions where the source material can be supplied to the e-beam location without the need for e-beam sputtering. For the system being examined here, the most relevant parameter that can be easily controlled is the substrate temperature.

To this end, the temperature of the sample was increased in steps to 900 °C, stopping momentarily to verify the continued presence of the Sn nanoparticles at 600, 700, and 800 °C. At 900 °C the sample was examined more closely. Figure 4(a) shows an image of a step edge harboring many Sn atoms. Figure 4(b) shows the same location, acquired five minutes later. The spontaneous dispersion of Sn atoms along the step edge and the formation of multiple Sn nanoparticles suggests that the Sn atoms are highly mobile.

One potential interpretation is that Sn atoms are rapidly diffusing across the graphene surface (too rapidly to image) but are stabilized at the step edge and become imageable. This interpretation suggests that a vacancy created anywhere in the graphene should result in the attachment of a randomly diffusing Sn atom. To explore this hypothesis, we used the e-beam to scan a small (~0.5 nm square) area over a random patch of the graphene. After a few seconds, instead of observing dopant insertion, a hole was created in the graphene indicating that the supply of Sn atoms in that location is lower than is present at the step edge. The spatial distribution of the supply of Sn atoms deserves further theoretical investigation, however our goal here was to empirically find conditions favorable for direct-writing.

Since there seemed to be an ample supply of Sn atoms at the step edge, a different strategy was employed. A small (~0.5 nm) subscan box was defined and positioned at the step edge, which provides a small, rapidly updating image of the irradiated location for real-time monitoring. As the step edge harboring Sn atoms was irradiated, Sn atoms were observed to attach to the underlying graphene. The subscan box was then manually repositioned to expose more of the pristine graphene. In this way, Sn atoms were drawn from the step edge and attached to vacancies/defects in a continuous manner with a single pass. This procedure is illustrated visually in the schematic in Figure 4(c).

Figure 4(d) shows a HAADF image of the resulting Sn dopant chain drawn in the graphene lattice. Figure 4(e) shows a magnified view displayed as a smoothed surface plot and artificially colored. The subscan box was moved as indicated by the arrow. After image acquisition, the process was continued, as shown in Figure 4(f). This second line required about seven minutes to complete. Figure 4(g) shows a magnified surface plot of (f). This direct-write process meets both criteria mentioned previously: 1) it occurs in a continuous fashion without an iterated checking procedure, as is required for the "beam dragging" process shown in Figure 3(a)-(d), and 2) it could be performed at a greater distance from the atomic source reservoir. The limit to this process remains an open question. Because of the continuous nature of this process, it should be amenable to automatic computer control. The potential for automation is a critical point when considering possible applications for atomic scale fabrication.

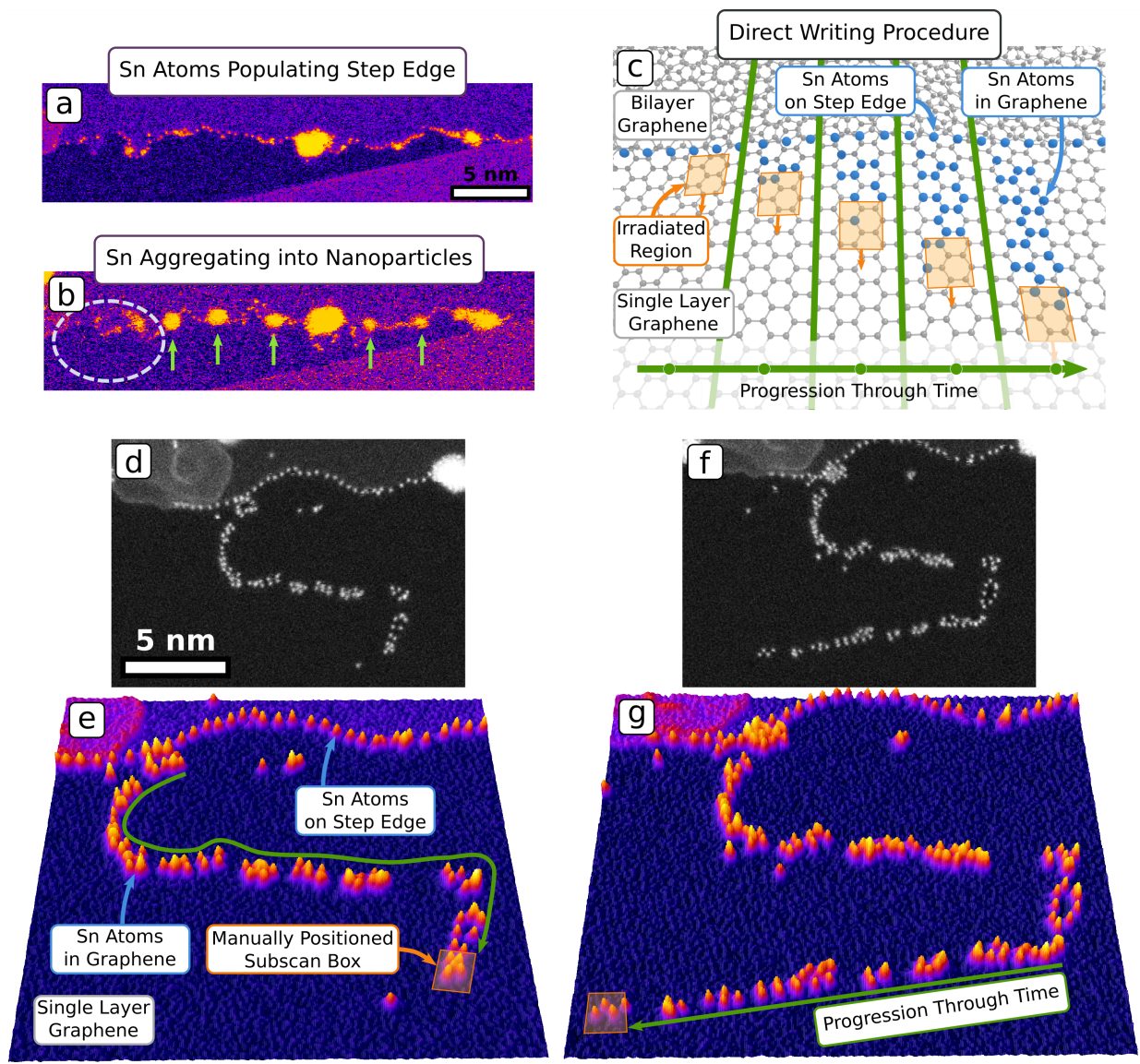

**Figure 4 Demonstrating atom-by-atom direct write using the e-beam at 900 °C.** (a) Initial state of a bilayer graphene step edge at 900 °C. (b) The same region shown in (a) acquired five minutes later. Additional Sn nanoparticles attached to the step edge indicate a high degree of Sn mobility along the edge, arrowed. Images shown in the (d)-(g) were acquired near the dashed region on the left. (c) Schematic illustration of the direct writing procedure. (d) HAADF image of a line of Sn dopant atoms directly written in graphene. (e) Enlarged and labeled surface plot of the image in (d). (f) The direct write process was continued, starting where the previous writing was halted. (g) Surface plot of the image shown in (f). Images in (a) and (b) were artificially colored using the "Fire" look up table in Imagej. Images (e) and (g) were artificially colored and smoothed to reduce noise.

It should be noted that we observe some Sn dopants that attach to the graphene surface outside the direct-write region. The most likely interpretation is that thermally induced vacancy diffusion enables vacancy migration from the irradiated region into the pristine graphene[52] after which Sn atoms can bond to the vacancy site. This evidence of Sn atoms diffusing across the graphene surface combined with the result of forming holes when direct deposition was attempted at random locations on the graphene suggests that the ratio of the defect generation rate to the adatom diffusivity (or concentration) plays a significant role in determining the outcome. These observations point toward fruitful avenues of inquiry for future experiments.

Clearly there are a variety of factors that need to be further understood and controlled to optimize atom-by-atom direct writing. Previous work in this context highlights the role of mobile carbon adatoms that spontaneously heal vacancies and larger structural defects.[53,54] At the same time, rapidly diffusing vacancies can markedly change the behavior of graphene at high temperatures.[52] E-beam patterning of corrals[24] can influence the influx of hydrocarbon contamination or provide a trap for long-range vacancy diffusion. The dynamics of these factors are clearly dependent on the sample temperature, and here we have demonstrated that temperature can also be used to influence the supply of source atoms. We have shown that a step-edge on the graphene and a deliberately created dopant chain provide local channels for the migration of source atoms and can enable continuous direct-writing of nanoscale features.

Both the imaging and direct-write process shown here were performed at the same microscope acceleration voltage (100 kV). The maximum energy transferred from the electron beam to a carbon atom in an elastic collision (where total kinetic energy is conserved) is above the knock-on threshold for graphene, which is critical for introducing the point defects that facilitate the attachment of the Sn atoms. However, this damage is detrimental for imaging, where the desire is to obtain sample information without changing the state of the sample. Ideally, the manipulation and imaging modes should be set above (e.g. >90 kV) and below (e.g. <80 kV) the knock-on threshold for graphene, which is around 85 kV.[40] However, changing the microscope accelerating voltage, while maintaining stability and resolution, remains a prohibitively slow process, despite some efforts at improvement.[55] What we employ here, instead of changing accelerating voltages, is changing the effective dose rate. The e-beam current was maintained at 17 pA, however, the dimensions of the scanning window determine how this current is spread or concentrated. For the overview images shown in Figure (d)-(g) we are applying a dose per unit area of $3.6 \times 10^6$ e/nm$^2$ and each image was completed in 8.7 s. Within the subscan region the calculation is more difficult because the subscan box is being moved across the sample in a dynamic fashion based on visual feedback about the sample state. The total dose required to draw the second line, shown in Figure 4 (f) and (g), was $4.2 \times 10^{10}$ e, however this dose was spread across the irradiated region sequentially and unevenly. For direct comparison to the imaging mode, we estimate the average dose per unit area to be $8.3 \times 10^9$ e/nm$^2$ during the direct-write process (dividing the total dose by the total irradiated area). This leads to an estimated atomic insertion rate of 8.5 s/atom. Given the ease and speed with which the dose rate may be changed based on pixel dwell time and scan area, this provides an accessible method for switching between imaging and manipulation modes that does not disturb the stability of the electron optical elements of the microscope.

Here, an e-beam directed, continuous atom-by-atom direct-write process on suspended single layer graphene has been demonstrated. This process relies upon the repeated ejection of C atoms from the substrate lattice as well as a continuous supply of source atoms. The generation of defect sites is governed by the e-beam accelerating voltage and beam current, here fixed at 100 kV and 17 pA. The incorporation of source atoms into the defect sites is governed by the rate of supply of source atoms to the generated vacancies. Substrate temperature, diffusion rates, and the local energetic landscape play significant roles in this process. We find that sample features, such as the graphene step edge, or artificially engineered features, such as the direct-write lines of atoms implanted into the graphene lattice, both serve a role in delivering source atoms to the defect sites. While the role of temperature might generally be expected, the role of local defects suggests that one could deliberately tailor the local environment to be favorable for the direct-write operation.

These findings indicate generally that e-beam atom-by-atom direct-write processes are possible provided that the source material can be supplied at the requisite rate. This suggests that a search for conditions which enable a direct-write process at arbitrary sample locations would be a fruitful direction of inquiry. Similar processes may enable atom-by-atom direct-writing on various other 2D materials or possibly inside 3D materials.

**Acknowledgment**

The authors would like to gratefully acknowledge the assistance of Zachary Gosser in the Sn evaporation process.

This work was supported by the U.S. Department of Energy, Office of Science, Basic Energy Sciences, Materials Sciences and Engineering Division (O.D. A.R.L., S.J.), and was performed at the Center for Nanophase Materials Sciences (CNMS), a U.S. Department of Energy, Office of Science User Facility.